\documentclass[5p]{elsarticle}
\usepackage{hyperref}

\journal{Journal of \LaTeX\ Templates}

\begin{document}

\begin{frontmatter}

\title{Reaction Losses of Charged Particles  in CsI(Tl) crystals}
\tnotetext[mytitlenote]{Fully documented templates are available in the elsarticle package on \href{http://www.ctan.org/tex-archive/macros/latex/contrib/elsarticle}{CTAN}.}

\author[add1,add2]{S. Sweany
\corref{mycorrespondingauthor}}
\author[add1,add2]{W.G Lynch\corref{mycorrespondingauthor}}
\author[add1]{K. Brown}
\author[add1,add2]{A. Anthony}
\author[add3]{Z. Chajecki}
\author[add6,add7]{D. Dell'Aquila}
\author[add4,add5]{P. Morfouace}
\author[add1,add2]{F. C. E. Teh}
\author[add1,add2]{C. Y. Tsang}
\author[add1,add2]{M.B. Tsang}
\author[add1]{R.S. Wang}
\author{and}
\author[add1,add2]{K. Zhu}

\address[add1]{National Superconducting Cyclotron Laboratory, Michigan State University, East Lansing, MI 48824, USA}
\address[add2]{Department of Physics and Astronomy, Michigan State University, East Lansing, MI, 48824, USA}
\address[add3]{Department of Physics and Astronomy, Western Michigan University, Kalamazoo, MI, 49008, USA}
\address[add4]{CEA, DAM, DIF, F-91297 Arpajon, France}
\address[add5]{Universit\'{e} Paris-Saclay, CEA, Laboratoire Mati\`{e}re en Conditions Extr\^{e}mes, 91680 Bruy\`{e}res-le-Chat\^{e}l, France}
\address[add6]{Dipartimento di Chimica e Farmacia, Universit\`{a} degli Studi di Sassari, Sassari, Italy}
\address[add7]{INFN-Laboratori Nazionali del Sud, Catania, Italy}

\begin{abstract}
To efficiently detect energetic light charged particles, it is common to use arrays of energy-loss telescopes involving two or more layers of detection media. As the energy of the particles increases, thicker layers are usually needed. However, carrying out measurements with thick-telescopes may require corrections for the losses due to nuclear reactions induced by the incident particles on nuclei within the detector and for the scattering of incident particles out of the detector, without depositing their full energy in the active material. In this paper, we develop a method for measuring such corrections and determine the reaction and out-scattering losses for data measured with the silicon-CsI(Tl) telescopes of the newly developed HiRA10 array. The extracted efficiencies are in good agreement with model predictions using the GEANT4 reaction loss algorithm for Z=1 and Z=2 isotopes. After correcting for the HiRA10 geometry, a general function that describes the efficiencies from the reaction loss in CsI(Tl) crystals as a function of range is obtained.

\end{abstract}


\end{frontmatter}


\section{Introduction}

Energy loss telescopes are often the tool of choice for detecting light charged particles and intermediate mass fragments emitted in nuclear collisions. At low energies, the detection material is frequently silicon. At intermediate energies above  10 MeV/A, a telescope more often consists of a Si detector backed by a thick scintillation detector. Thallium doped Cesium Iodide crystals (CsI(Tl)) are often selected to be this last scintillation detector. CsI(Tl) crystals have excellent energy resolution, are easily machinable,  not especially hygroscopic, and produce light at wavelengths that match well with the response of silicon based photo-diodes \cite{Gong1988,Wagner2001}. They are also relatively inexpensive, allowing large arrays to be built. 

The thicknesses of such scintillators are set by the range of the most penetrating particle to be measured. As the particle energy increases,  the required thickness of such scintillators increases, and the likelihood that the incident particle will suffer a nuclear interaction within the scintillator also increases. In such cases, one observes events in which the incident particle only deposits a fraction of the expected electronic energy loss in the scintillator, reducing the electronic signal from the scintillator to a fraction of its normal value. The correlation between the energy loss "dE" in the first detector and the energy "E" in the last detector for this particle becomes distorted, misplaced from  the characteristic Particle IDentification (PID) line corresponding to that particle. For illustration, we use the HiRA10 Si-CsI array that is described in the next section. However, the reaction loss function derived at the conclusion of this paper is independent of detector geometry, allowing for the easy estimation of reaction loss probabilities.

\begin{figure}
\centering
\includegraphics[width=\columnwidth]{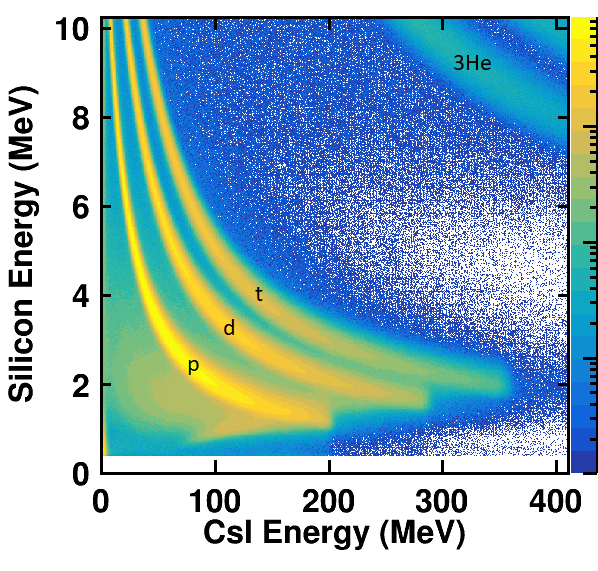}
\caption{Going from left to right the PID lines shown are protons, deuterons and tritons. The background haze to the left of the lines comes from reaction loss and out-scattering events.}
\label{HiRAPID}
\end{figure}

Figure \ref{HiRAPID} shows a typical HiRA10 PID plot that displays the characteristic PID lines for protons ($p$), deuterons ($d$) and tritons ($t$) appearing as hyperbolic curves. Here, heliums ($^{3,4}He$) lie slightly outside the boundaries of the figure. An intensity scale is provided at the right of the figure. Particles that suffer  reaction losses appear in the lower intensity haze to the left of the PID lines. In Ref. ~\cite{MORFOUACE201745} it was found that reaction  and  out-scattering losses can be as high as 40\% for hydrogen and helium isotopes penetrating through 10cm CsI(Tl) crystals. Since the out-scattering loss is largely due to the Coulomb interaction, its contribution is smaller. The major contribution and therefore the biggest uncertainty in the correction for these losses is the reaction loss component, i.e. the reduction of the energy deposited in the active detector volume through electronic stopping due to the occurrence of a nuclear reaction with the detector material.

The purpose of this work is to use experimental data obtained with the HiRA10 dE-E telescopes \cite{DELLAQUILA2019162} to determine the reaction losses and check the reaction loss models studied in the work of Morfouace et. al. ~\cite{MORFOUACE201745}. Ideally, such studies should be performed at a facility capable of producing monoenergetic beams of light charged particles which would be used to illuminate the CsI(Tl) crystals. Unfortunately, there are very few accelerators that can provide such beams, making it difficult to obtain the  beam times that will be required for such measurements. 

This motivates the present approach  of determining reaction losses from the same experimental data for which one needs the reaction loss correction. To apply this approach we need Geant4 simulations that use reasonable descriptions of the reaction cross sections for the incident particle on the nuclei that comprise the detector. Accordingly, we performed such simulations using the NPTool framework \cite{Matta2016}. These calculations simulated light charged particle spectra for protons, deuterons, tritons, $^{3}He$, and alpha-particles. Using the simulated and experimental data,  we expand upon methods developed by Avdeichikov et.al. and Siwek et. al. \cite{Avdeichikov1999,Siwek2002} to extract the reaction losses from the HiRA10 charged particle data in CsI(Tl) crystals and compare them to the parameterizations of  Morfouace ~\cite{MORFOUACE201745}. 

In the following section of this article we discuss the detector system used in this study, known as HiRA10. Section 3 delves into the methods developed for extracting the reaction losses through the comparison of simulated and experimental data. In section 4 we present our conclusions. 

\section{HiRA10}

The data for this study were collected using the HiRA10 Array, which consists of 12 dE-E charged-particle telescopes. Depending on the needs of a given experiment, the telescopes can be configured into different geometries. In the present experiment, the 12 telescopes were arranged into 3 towers with 4 telescopes in each tower. The telescopes cover a large range of polar angles. While the reactions of incident particles in each telescope are fundamentally the same, the relative intensities and energies of particle species varied throughout the array and differed for different projectile energies. 

\begin{figure}
\centering
\includegraphics[width=\columnwidth]{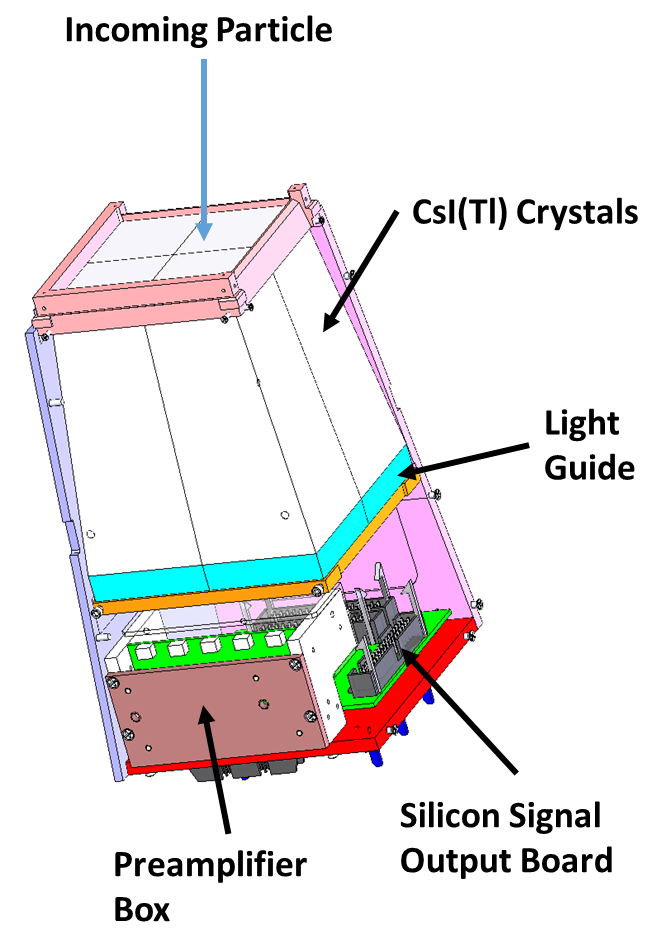}
\caption{HiRA10 Telescope mechanical drawing is shown with the silicon detector and two aluminum side panels removed. Four CsI(Tl) crystals and their four light guides can be seen in the figure. Also shown are the preamps for the CsI(Tl) signals and two connectors used to transmit the signals out of the aluminum box that encloses the telescope. }
\label{HiRAPic}
\end{figure}

The first component of each HiRA10 telescope in the array is a 1.5 mm Double-Sided silicon Strip Detector (DSSD). Each DSSD used in this experiment had 32 1.95 mm wide front and 32 1.95 mm wide back strips oriented perpendicular to each other, allowing for the position where a particle enters the detector to be determined. The silicon detectors used in the HiRA10 Array are identical in design to those used in its predecessor, the High Resolution Array (HiRA) \cite{Wallace2007}. Backing the DSSD is a pack of four 10 cm long CsI(Tl) crystals supplied by Scionix in the Netherlands. Each crystal was cut so that they taper from the front of the telescope, (35x35mm) to the back (44.6x44.6 mm). The front and back faces of the crystals are polished to a machined finish while the sides were sanded with 240 grit sand paper along the axis running perpendicular to the entrance window. The uniformly diffuse reflecting surface provided by sanding the sides of the crystals is important to achieving a position independent light output throughout each crystal \cite{Gong1988}. A plastic light guide was attached to the back of each crystal using GE Bayer RTV615 silicon rubber glue. A 18x18 mm$^{2}$ Hammamatsu S3584-08 photo diode was attached to the back of the light guide using the same RTV615 glue. Each crystal was wrapped using a proprietary wrapping method developed by Scionix, wherein the sides were wrapped with a diffuse reflecting foil and the entrance window of the crystal was covered with a layer of 0.29 $mg/cm^2$ double-sided aluminized mylar foil. The light output uniformity of each crystals is determined by scanning it with an $^{241}Am$ alpha source collimated to a spot size of 3 mm over a 3x3 grid \cite{SeanThesis}. The light output of each crystal did not vary by more than 1\% for the 9 points measured during each scan.  Details on the testing and performance of the silicon detectors can be found in \cite{Wallace2007,vanGoethem2004}. We show a mechanical drawing of a single telescope in Figure \ref{HiRAPic}.

Charged particle data were measured with the HiRA10 array during a recent experimental campaign at the National Superconducting Cyclotron Laboratory in East Lansing, Michigan, using the Coupled Cyclotron Facility. Here we present data for three different reactions: $^{48}Ca+^{124}Sn$, $^{40}Ca+^{112}Sn$, and $^{48}Ca+^{64}Ni$ that were measured at beam energies of 140 MeV/A. For this analysis data from two of the detectors within the middle tower was used. This was due to the combination of reasonably high statistics for reaction losses as compared to the most backward angle telescopes and a limited amount of punch-through background as compared to the most forward angle telescopes.

\section{CsI Reaction Loss Analysis}
\begin{figure}
\centering
\includegraphics[width=\columnwidth]{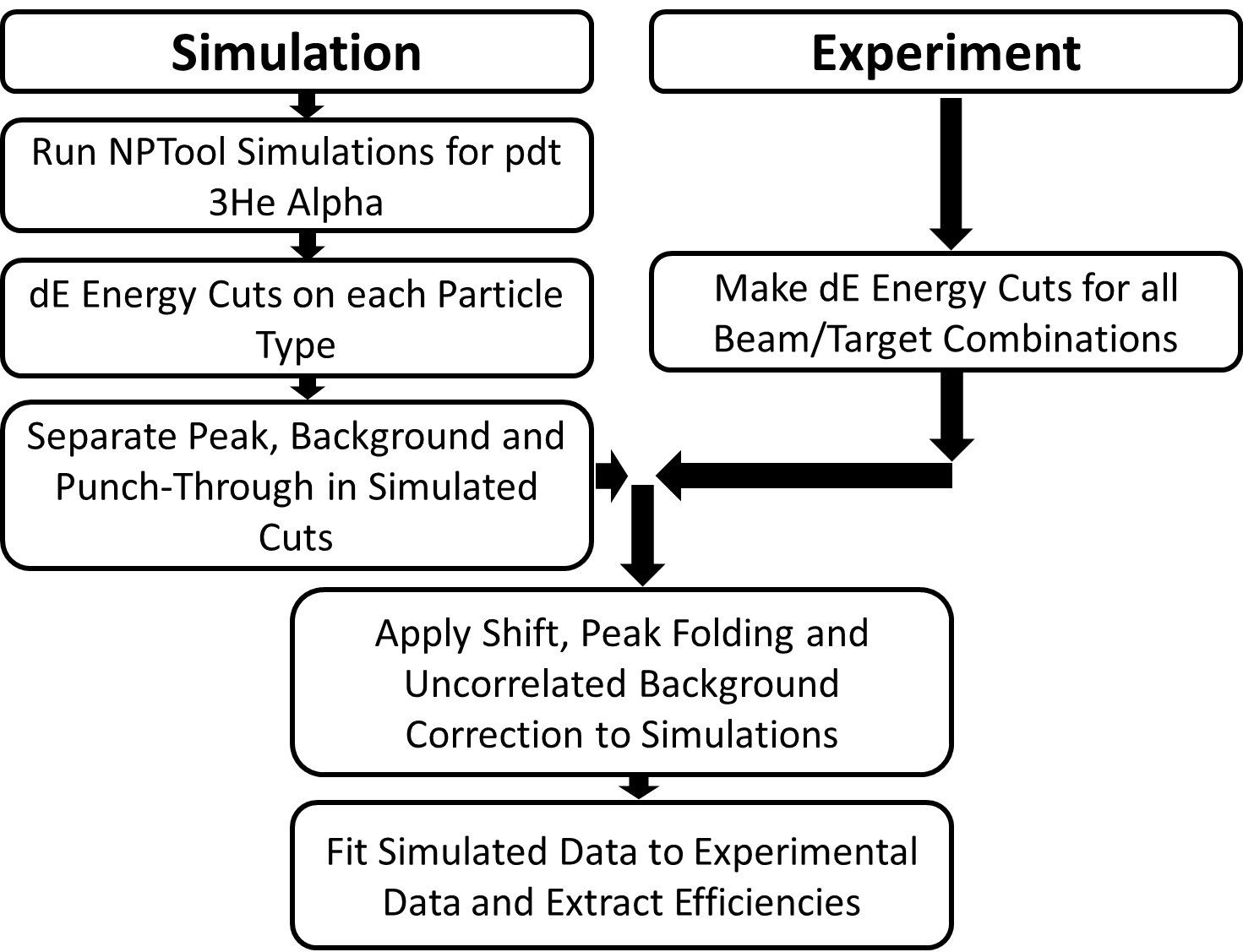}
\caption{Reaction Loss Analysis Flow Chart}
\label{FlowChart}
\end{figure}

Our reaction loss analysis proceeds as follows. First, we set a gate on position in the silicon detector to remove events in which particles are injected within 2 mm near the edge of a crystal and are more likely to be scattered out of the crystal due to Coulomb scattering by the Cesium (Cs) or Iodine(I) nuclei in the detector. Applying this gate, we performed Geant4 simulations to create events with and without nuclear reactions between the incident light particles and the Cs or I nuclei in the crystals. We impose  gates on the energy deposited in the silicon dE on both the simulated and measured experimental events. We then separate the simulated events into 1) good (non-reacted) events where the incident particles lose their full incident energy in the telescope by electronic stopping and 2) (reacted) events in which a reaction or out-scattering occurs. Next, corrections are made to the simulated data to account for small differences in the resolutions or calibrations of the experimental and simulated data. Finally, the simulated  histograms are used to create fitting functions, which are then fitted to the corresponding dE cuts in the experimental data. Using these scaled histograms, the detection efficiency is then extracted for each of the light charged particle species. A flowchart describing the analysis method is shown in Figure \ref{FlowChart}.

\subsection{Simulations}
The first step of the reaction loss analysis is to simulate individual spectra for  p, d, t, $^{3}He$ and $^{4}He$ with Geant4 simulations (version 10.04.02) using energy spectra similar to those of the corresponding particles in the experimental data. These simulations were performed using the NPTool framework with the standard Geant4 EMPhysicList Option4 to simulate electromagnetic processes \cite{Matta2016}.  Different cross section parameterizations are available in GEANT4 to simulate the reactions in CsI(Tl) crystals for p,d,t,$^{3}He$ and alpha particles. Following Ref. ~\cite{MORFOUACE201745}, we employ the Shen parameterization \cite{Shen1988} to describe the reactions for d,t,$^{3}He$ and alpha particles and the Grichine parameterization \cite{Grichine2009} for protons. We did not use Tripathi or the INCL parameterization since both produce  reaction losses that are similar to the Shen parameterization \cite{MORFOUACE201745}. 

As expected, the number of charged particles lost due to reactions and out-scattering in the CsI(Tl) increases with the range of the  particle, which is governed by its initial energy. Since the energy loss in the dE decreases inversely with energy, we can probe this range dependence by setting cuts on the dE. By selecting lower values of dE for a given isotope, we select events with higher incident energies and longer ranges and larger associated reaction losses. For each dE gate, we calculate CsI(Tl) spectra for the proton, deuteron, triton, $^{3}He$ and alpha particles that are consistent with that gate. Since the dE value varies inversely with the E value for a given isotope, it is natural that the width of the dE cut must decrease with dE in order to make an effective selection on the incoming energy. Accordingly, we impose narrow dE cuts on the order of 25 KeV wide for dE $<$ 1.5 MeV, 50 KeV for 1.5MeV $<$ dE $<$ 3.5 MeV and  100KeV for dE $>$ 3.5 MeV. Subject to these cuts on the energy loss in the silicon, dE, we create experimental and simulated spectra of the energies deposited in the CsI(Tl) detectors. 

\begin{figure}
\centering
\includegraphics[width=\columnwidth]{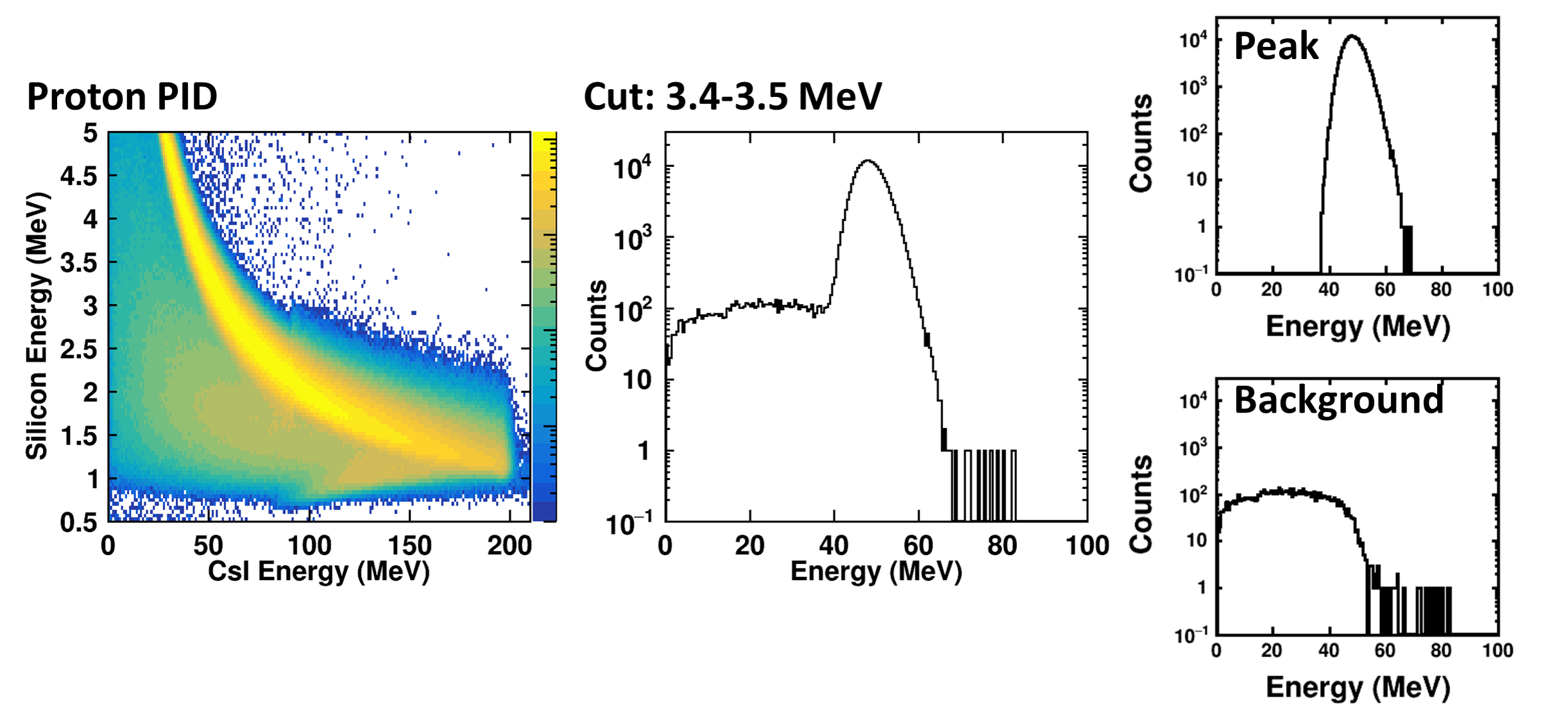}
\caption{Left: Simulated proton PID spectra. Middle: Example of a CsI(Tl) E spectrum for a dE cut from 3.4-3.5 MeV. Right: Simulated histograms after separation into "reacted" and "unreacted" protons: top shows protons which deposited all of their energy and the bottom shows protons that didn't deposit all of their energy due to either out-scattering or a reaction loss event.}
\label{dECut}
\end{figure}

We classify the simulated CsI(Tl) events into reacted, non-reacted or punch-through events. (In the latter events, the particle penetrates fully through the CsI(Tl).) For each dE cut, three CsI(Tl) spectra are created for: 1) particles (p, d, t, $^{3}He$ or alpha) that lose their total energy in the CsI(Tl) crystal, 2) particles that deposit part of their energy in the CsI(Tl) detector before reacting or scattering out of the detector, and 3) particles that traverse fully through the detector. The last group is referred to as “punch-through” particles which are identified as events with an initial energy above the punch-through point. We label events as reaction loss/out-scattering events when they do not punch-through the CsI(Tl) detector but nevertheless still lose more than 1.5\% of its initial energy. In between these two thresholds exists a ~5-10 MeV region on either side of the punch-through point where range straggling can cause mixing between normal events, reaction loss events, out scattering events and punch-through events. For simplicity, we do not extract reaction losses using data in these regions where the mixing of processes is unclear. An example showing the separation between full energy deposition and reaction loss/ out-scattering events at energies well below the "punch-through" energy is illustrated in Figure \ref{dECut} for a 100 KeV wide dE cut centered at 3.45 MeV. For this cut, the reaction loss contribution greatly exceeds the out-scatting contribution.

\subsection{Corrections}
Several corrections are made to the simulated data before they can be compared to the experimental data. The first is to apply a shift to the simulated data to match the peak positions in the experimental data. This is necessary because the nonlinear isotope dependence of the kinetic energy vs. light output relationship has to be taken into account in the simulation so that it matches the experimental spectra in Figure \ref{HiRAPID}. This is done by simply finding the position of relevant particle peaks (d, t, $^{3}He$ and alphas) in both simulation and experimental data and dividing the position of the peaks to obtain a shift factor. This factor is then multiplied by the energy for each of the simulated reacted and non-reacted events, which shifts the simulated peaks and background to their appropriate locations in the PID spectrum. 

At higher energies, corresponding to lower dE cuts, it becomes necessary to take punch-through events into account. The few deuteron and tritons with sufficient energy to punch through the detector contribute insignificantly in the current analysis.  Proton punch through events, on the other hand, are more common and where they appear at high energies in the CsI(Tl) these punch-through events account for the majority of the background. For energies greater than 200 (198.5 to be exact) MeV, the proton punch-through tail folds over the proton PID line and trends towards lower energies in Figure \ref{HiRAPID}. We make some adjustments to match the calculated energy deposition for simulated punch-through events, to the measured ones. 

Another problem with correcting for the punch-through is that we do not have information about the proton spectra above the punch through energy of 198.5 MeV. At such high energies, the precision of energy loss in the silicon detector is insufficiently precise to allow us to constrain the proton energy spectrum. We therefore assume for this region that reaction cross sections of the incident particles on the Cs or I nuclei do not change significantly at energies near and above 200 MeV. 

To achieve accurate fits to the measured spectra, we find it necessary to modify the resolution of the simulated data corresponding to full energy deposition in the crystal. The peaks in the experimental data are somewhat broader and display a longer high energy tail than the simulated data. These differences in peak shape have two causes: 1) the actual peak widths in the CsI(Tl) crystal are broader with a tail extending to higher pulse heights that is not replicated by the simulated data and 2) the full energy peak for the stopped charged particle is sometimes modified by coincident summing of an additional signal from an uncorrelated $\gamma$-ray or electron in the same detector in the same event. 

Regardless of the origin, the difference between the measured and calculated peak shape must be addressed. If the peak width of the simulations is narrower than that in the data, it will cause the fitting procedure to increase the background from reactions in order to fit the data near the peak, causing a nonphysical enhancement to the number of reacted particles. If the simulated  peak width is wider than the data, it is also problematic because the fit will favor decreasing the number of reacted particles to better reproduce the data near the peak. 

 We have studied the distortions of the full energy peak due to the uncorrelated background. We find that much of the uncorrelated background in a dE-E detector mainly results from the coincident summing involving one charged particle and one neutral particle. When more than one particle hits a given CsI(Tl) crystal, pile-up occurs and the resulting larger energy is incorrect and may push the particle outside its PID line. Coincident summing of two charged particles is suppressed by the granularity of the silicon detector. Each CsI(Tl) crystal is covered by 256 pixels in the Si detector.  Most of the events with two charged particles in a single CsI(Tl) hit different silicon strips and can thereby be identified and rejected so that the numbers of pile up events involving two charged particles are correspondingly reduced.

We correct for the uncorrelated background by taking such pile-up events into account. The number and characteristic properties of such pile up events is  determined by looking for events within a telescope where there is a good event (Silicon Hit + CsI Hit) and an uncorrelated event (CsI hit + No Silicon Hit) in the one CsI(Tl) that is not adjacent to the first CsI(Tl) but rather lies diagonally across from it within the telescope can. This gives us an estimate of the fraction of the events that are affected by this form of uncorrelated background.

\begin{figure}
\centering
\includegraphics[width=\columnwidth]{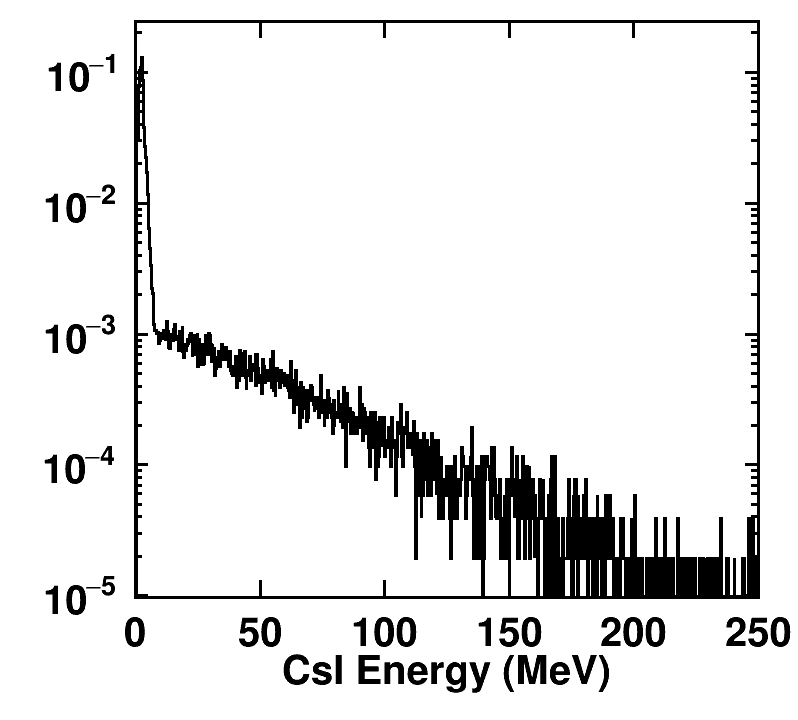}
\caption{Uncorrelated background energy distribution in the CsI normalized to 1.}
\label{UncorrEDist}
\end{figure}

After estimating what fraction of events are affected, the next step is to determine how this affects the spectrum. To do this, the energy distribution of the pure uncorrelated background events are extracted by looking for events with 1) a hit in a silicon detector without corresponding signal in the CsI(Tl) behind it and 2) a neutral particle detected in the one CsI(Tl) that is not adjacent to the first CsI(Tl) but rather lies diagonally across from it within the telescope can. This situation is analogous to the case of a purely uncorrelated event where a low energy particle hits the silicon while a neutral particle hits the CsI(Tl) behind it. The extracted neutral particle pulse height distribution  is shown in Figure \ref{UncorrEDist} after employing the energy calibration to convert these pulse to their corresponding charged particle "energy"  values. The CsI(Tl) energy distribution for uncorrelated background is then used as a generating function to add a pile-up correction to the simulated data. The above case accounts for about ~80\% of the uncorrelated events. We performed additional studies to account for the situations where there were 2 and 3 uncorrelated particles in the crystal as well as corrections for the uncorrelated particles scattering between crystals. 

This correction for the pile-up does not strongly affect the reaction loss of the particle involved in the pile-up. Instead it influences more strongly the reaction loss background for heavier $^{3}He$ and alpha particles. Uncorrelated events piling up on p,d,t data appear to the high energy (right) side of the p,d or t peaks where they can be on the same order as reaction loss events involving the helium isotopes. Without the pile-up correction we find that the reaction losses from $^{3,4}He$ will be overestimated. 

\begin{figure}
\centering
\includegraphics[width=\columnwidth]{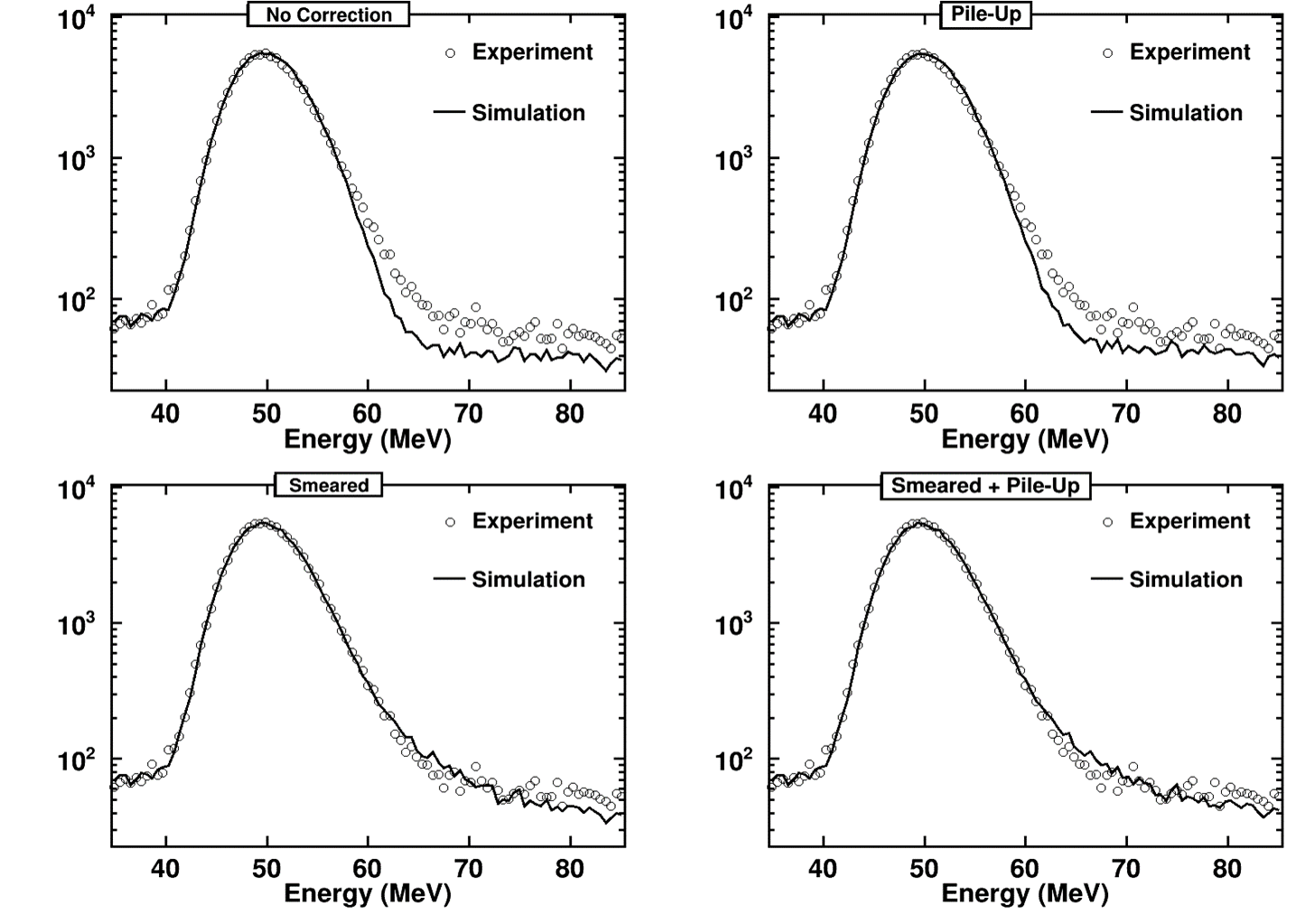}
\caption{The four panels show protons for a cut on dE from 3.4-3.5 MeV. Open points show the experimental data and the solid line shows the simulated protons with different corrections applied as follows for the different panels. Upper Left: No correction, Upper Right: Correction for uncorrelated background, Lower Left: Correction for Gaussian folding, Lower Right: Both corrections.}
\label{CorrectionExample}
\end{figure}

We now return to the small difference between the calculated and measured line shape for properly identified protons, deuterons, and tritons. The main effect here is a high energy tail on the CsI(Tl) spectrum that represents a few percent of the stopped events in the peak. The origin of this effect is not clear, but it does not reflect the intrinsic resolution of the CsI(Tl), which is of the order of 1\%. It could reflect rare anomalous fluctuations in the dE energy loss signal caused by channeling or by variations in the thickness of the silicon detectors, for example. To correct the calculated line-shape, we randomly add "noise" according to a noise distribution to each simulated CsI(Tl) signal. This noise spectrum is calculated using  a sum of "modified Gaussians" that are products of an exponential and the complementary error function, following an approach similar to that of ref. \cite{Brotels1987,Pomme2015}. We use a linear combination of three "modified Gaussians" with each having different exponential tails, but retaining the same standard deviation and mean. We adjust the tail parameters of these modified Gaussians to replicate the shape of the peak; two of the modified Gaussians have a tail on the right side of the peak and the third has a tail on the left side of the peak. We also adjust the probabilities for each of these functions to optimize the description. Figure \ref{CorrectionExample} shows protons within a dE cut from 3.4-3.5 MeV with the panels showing both the experimental data (open points) and simulated data (solid lines) with different corrections being applied. The upper left panel shows the differences between experimental and simulated line shapes with no correction. Adding pile up corrections (upper right panel) changes the simulated line shape very little. The bottom left panel shows how the Gaussian smearing adjusts the high energy tail of the simulated spectrum to better approximate the data. Again, adding pile-up corrections (lower right panel) changes the overall line shape insignificantly.

\subsection{Fitting}
After corrections are made to the calculated spectra for all of the simulated effects involving the reacted and non-reacted particles, these histograms are then fit to the observed spectra to determine the reaction losses in the crystals. This process is repeated for each of the individual dE cuts.

The calculated peak and background histograms for different light charged particle species are parameterized for fitting using Equation \ref{ReactionLossScaling} where $P_{i}$ and $B_{i}$ represent the peak and background histograms for particle i within a dE gate, respectively. $A_{i}$ is an overall normalization constant that reflects the number of incident particles at an incident energy corresponding to the energy loss dE gate for particle i. $A_{i}$ adjusts the overall magnitude of the simulated histograms. A second normalization constant, ${a}_{i}$ (typically of order unity) is applied to the background histogram to account for differences in the predicted and observed reaction loss background. A value of ${a}_{i}=1$ would imply that the reaction loss corrections of Ref. ~\cite{MORFOUACE201745} perfectly describe the data for this particle and at the energy being calculated. 

\begin{equation}
f(dE) = \Sigma{^n_i}A_{i}(P_{i} + {a}_{i}B_{i})
\label{ReactionLossScaling}
\end{equation} 

To check the consistency of the reaction loss determination, cuts for $^{48}Ca + ^{124}Sn$, $^{40}Ca + ^{112}Sn$, and $^{48}Ca + ^{64}Ni$ were simultaneously fitted using the simulated data. By choosing three systems with different N/Z ratios, we gain sensitivity that helps us to constrain the reaction losses from the different isotopes. During fitting, the overall scaling parameter ${a}_{i}$ for the different particle species was allowed to vary freely, reflecting the fact that the relative abundances of the heavier hydrogen and helium isotopes become larger as the projectile and target become more neutron rich.  However, the same background scaling factor ${a}_{i}$ is used for all systems, because it is proportional to the ratios of the measured/calculated reaction losses which should be independent of the reaction. Thus, the simultaneous fitting of 3 systems allows the values for ${a}_{i}$ to be better constrained, which is the main goal of this work. For higher energy hydrogen isotopes that penetrate the last 5 cm of the CsI crystal, we obtained average values for ${a}_{i}$ of approximately 1.0 for protons, 1.24 for deuterons and 1.19 for tritons. For the helium isotopes, the average of best fit values for ${a}_{i}$ remained close to 1.0 over the energy range we could measure.   

\begin{figure}
\centering
\includegraphics[width=\columnwidth]{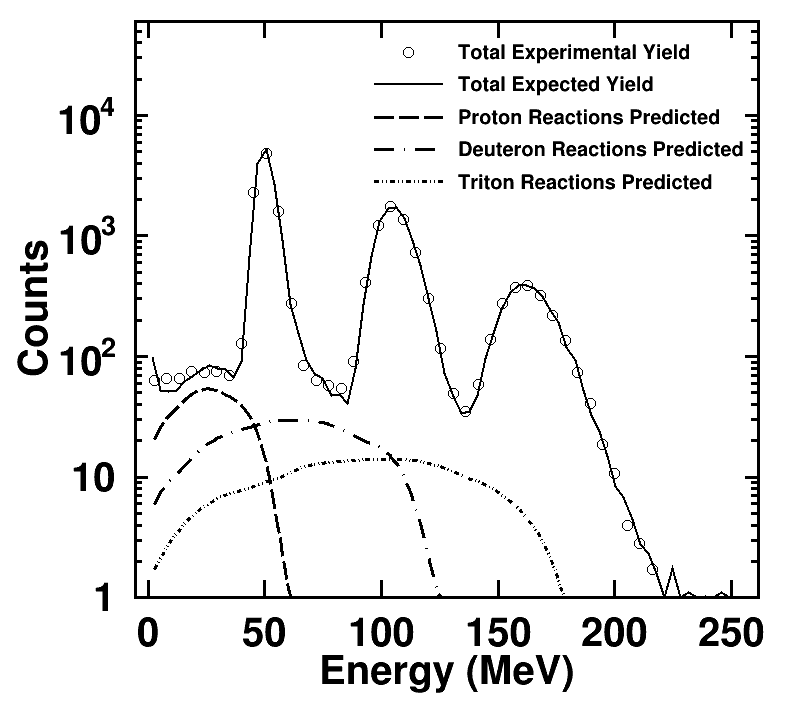}
\caption{Fit on a dE cut between 3.4 and 3.5 MeV for $^{48}Ca+^{124}Sn$. Experimental data are shown as the small dots while the overall fit using the simulated histograms are illustrated with the large dashes. Fits to the background are shown for protons, deuterons and tritons and shown as solid lines of varying thickness.}
\label{PDTCut}
\end{figure}

Figure \ref{PDTCut} illustrates the quality of the reproduction of the measured data by the simulation for a dE cut centered on 3.45 MeV. Here we show the experimental yield as well as the simulated background spectra for protons, deuterons and tritons that react within the CsI(Tl) crystal. From the fits described above, reaction losses and resulting efficiencies for the five light particles are extracted by dividing the total number of counts in the scaled peak by the sum of the counts in the scaled peak plus background. While we are able to determine the reactions losses up to the punch through energies for Hydrogen isotopes in the HiRA10 crystals, efficiencies for $^{3}He$ and alpha parties, could only be extracted up to 375 and 500 MeV, respectively, due to the lack of statistics at higher energies. 

\begin{figure}
\centering
\includegraphics[width=\columnwidth]{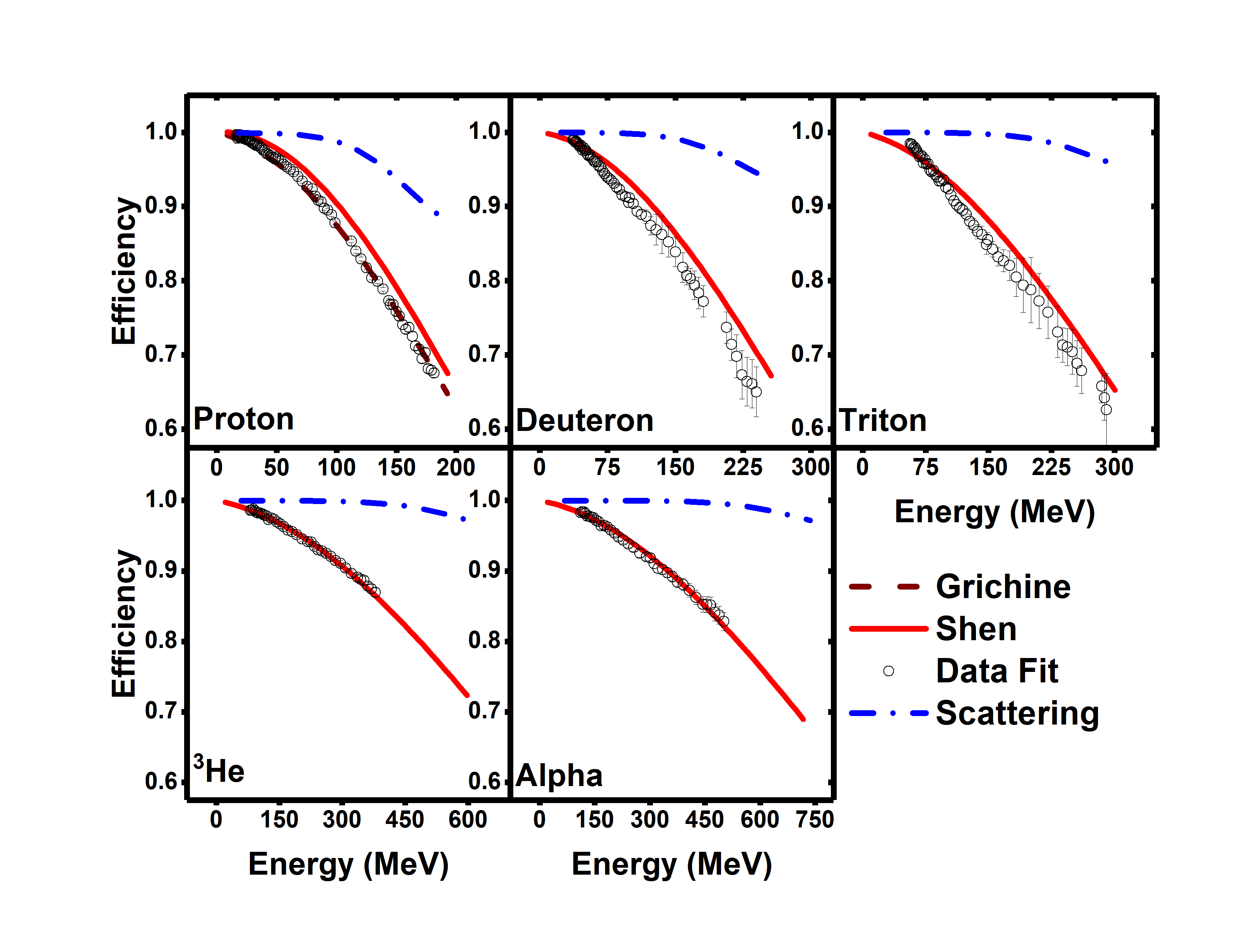}
\caption{Light charged particle efficiencies as a function of energy. Each panel shows the extracted efficiency (open points) as well as the simulated efficiency using the Shen cross section paramterization (solid lines) and the efficiency coming from only multiple scattering (dot dashed lines). For protons, we also include the efficiency using the Grichine parameterization (long dashed line), which describes the reaction losses better.}
\label{EnergyEff}
\end{figure}

Figure ~\ref{EnergyEff} shows the efficiency curves for our data. Error bars for the extracted efficiencies (data points) reflect both statistical errors as well as ambiguities in fit parameters from the scaled peak and background histograms. Such systematic uncertainties influence the uncertainties for deuterons and tritons most because the values extracted for their reaction loss contributions are somewhat anti-correlated, where an anomalously large value for the scaling parameter for the deuteron background was usually accompanied by an anomalously small value for the triton background and vice versa. 

The extracted efficiencies in Figure ~\ref{EnergyEff} reflect both reaction and out-scattering losses so the total loss is dependent on the geometry of the detector and any collimation that restricts the volume of the crystal used for detection. The largest out-scattering contributions (dot-dashed curves) occur for protons, due to its larger charge/mass ratio. They are on the order of 10\% of the total losses at the highest energy (200 MeV). In this work, we require particles to pass through a region of the silicon detector that lies at least 2 mm from the edge of the CsI(Tl) crystal. That requirement was not imposed in ref.  \cite{MORFOUACE201745}.  Vetoing charged particles that pass through these outer silicon strips reduces the numbers of particles lost to out-scattering and the efficiencies in this work are somewhat larger that those in \cite{MORFOUACE201745} but still within their stated 3\% uncertainty for proton and He isotopes.

Since the measured and calculated reaction losses for $^{3,4}He$ are in excellent agreement with the available data, it appears likely that the Shen parameterization  describes the reaction losses accurately at energies beyond where we can confirm their accuracy. For protons, the Grichine parameterization (black dashed line) predictions agree with the data better than the Shen parameterization. The simulated efficiencies for deuterons and tritons overestimate the measured values by about 5\% at the punch-through point. It appears that the Shen cross sections adopted by Morfouace et al. \cite{MORFOUACE201745} are smaller than what is needed to predict the reaction losses for deuterons and tritons correctly. This difference, however, is systematic and a simple empirical correction can be adopted.

\begin{figure}
\centering
\includegraphics[width=\columnwidth]{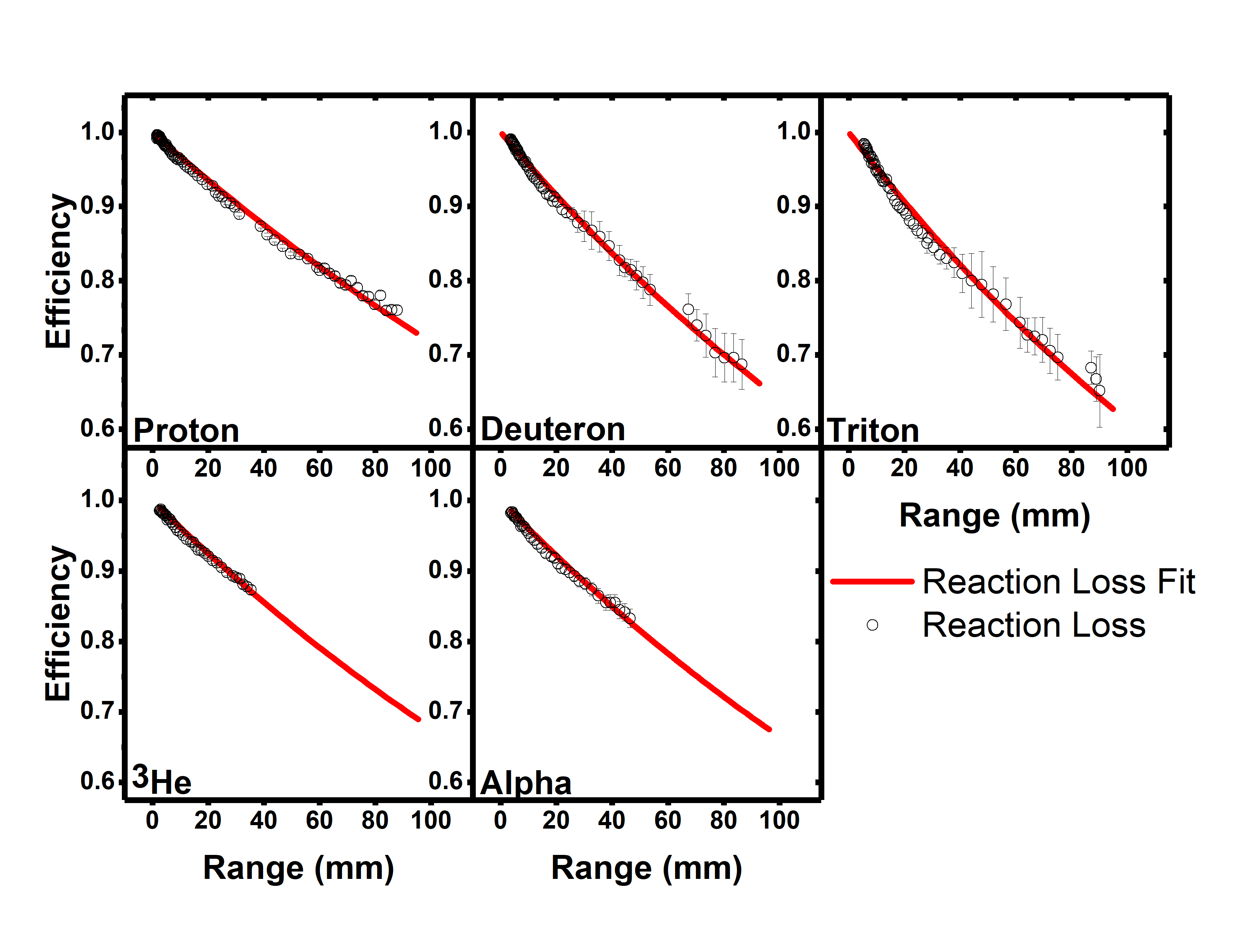}
\caption{Efficiency as a function of range for the five light charged particle species. The open points come from the fitted data with multiple scattering efficiency removed and the solid blue line is the fit of these points using Equation \ref{EffRangeFit_MultScat}. The fitting parameter "c" is $3.33*10^{-3}$ for protons, $4.45*10^{-3}$ for deuterons, $4.92*10^{-3}$ for tritons, $3.9*10^{-3}$ for $^{3}He$ and $4.08*10^{-3}$ for alphas.}
\label{RangePlot}
\end{figure}

Out-scattering primarily occurs for particles that enter the crystals near the crystal boundary. It becomes more prevalent for smaller crystals where a larger fraction of the incident particles can enter the crystal near the crystal boundary. Since GEANT4 does not allow calculations without Coulomb interactions, performing a "pure" reaction loss calculation is not an option. Out-scattering can be negligible for a highly collimated beam of particles that enters  a very large crystal, providing one method to get the "pure" reaction loss efficiency. Because out-scattering depends on crystal geometry, it is interesting to know what the current measurements imply in the limit of negligible out-scattering. For our smaller crystals, one can approximate the corresponding values for the effect of pure reaction losses by renormalizing away the fraction that out-scatters. We do this by dividing the full efficiency in Figure ~\ref{EnergyEff} by the fraction that does not out-scatter (blue dot-dashed curves in Figure ~\ref{EnergyEff}). This effectively renormalizes the flux in the full calculations so that there is no loss in flux due to out-scattering. 

So there is two approaches to calculate "pure" reaction loss efficiencies: 1) calculating it with a highly collimated beam and 2) dividing away the out-scattering efficiecy. We have compared the two approaches for protons, which have the largest out-scattering corrections and find that the two approaches agree to better than 1\%. This is useful because we can obtain pure reaction loss efficiencies by dividing the experimentally measured efficiency by the calculated out-scattering efficiency, which allows us to obtain "pure" reaction loss efficiencies for $d$ and $t$, which GEANT4 does not currently reproduce. 

The resulting "pure" reaction loss efficiency plotted as a function of range (the calculated depth of the detector that is penetrated by a particle) is shown in Figure \ref{RangePlot}. The energy to range conversion is performed with LISE \cite{BAZIN2002307}. Here, we plot the pure efficiency as a function of range instead of energy, because the incident particle flux decreases exponentially with the thickness of the detection material traversed in the limit of a constant reaction cross section.  For simplicity, we have fitted the "pure" reaction loss efficiency values in Figure \ref{RangePlot} with Equation \ref{EffRangeFit_MultScat}. This results in the  fits shown as solid red lines. The fitting parameters "c" are $3.33x10^{-3}$ for protons, $4.45x10^{-3}$ for deuterons, $4.92x10^{-3}$ for tritons, $3.9x10^{-3}$ for $^{3}He$ and $4.08x10^{-3}$ for alphas. It is then straight forward to apply the efficiency correction to other detector systems; all that is required is to simply multiply the efficiency for multiple scattering generated by GEANT4 by the reaction loss efficiency from Equation \ref{EffRangeFit_MultScat}. 

\begin{equation}
E_{Reactions}(R) = e^{-cR}
\label{EffRangeFit_MultScat}
\end{equation}

\section{Conclusions}
Using a technique that can be applied to different energy loss telescopes, we have carefully tested the recent reaction loss modeling of Ref \cite{MORFOUACE201745} for Hydrogen and Helium isotopes in CsI(Tl) experimentally. We find that our measured proton reaction losses match well with the Grichine cross section parameterizations at high energies. Similarly, $^{3}He$ and alphas also match very well with the Geant4 calculations with the Shen parameterization. On the other hand, both deuterons and tritons have lower efficiencies than the Shen parameterization predictions by several percent, varying from low to high energies. This is consistent with the finding of  Morfouace et. al \cite{MORFOUACE201745} who found that all of the available GEANT4 parameterizations underestimated the reaction cross section for deuterons on CsI(Tl). The difference, however, is relatively small and, once realized, can be easily corrected after running simulations with the Shen parameterization. We have also derived a formula with best fit parameters for the reaction loss efficiency as a function of range for CsI(Tl) crystals. This formula can be applied to any detector system by providing the GEANT4 efficiencies from multiple scattering which is specific to an individual detector.  
 
\section{Acknowledgement}
This work is supported by the U.S. National Science Foundation under Grant No. PHY-1565546 and the U.S. Department of Energy (Office of Science) under Grant No. DE-SC0014530, DE-NA0002923, and DE-NA0003908. The authors would also like to thank the HiRA group.

\bibliography{mybibfile}

\end{document}